\documentclass{pasj00}

\def\Porb{$P_{\rm orb}$}
\def\Psh{$P_{\rm SH}$}

\begin{document}
\SetRunningHead{D. Nogami et al.}{Photometry of DM Lyr}

\Received{2002/00/00}
\Accepted{2002/00/00}

\title{The SU UMa Nature of the Dwarf Nova, DM Lyrae}

\author{Daisaku \textsc{Nogami}}
\affil{Hida Observatory, Kyoto University,
       Kamitakara, Gifu 506-1314}
\email{nogami@kwasan.kyoto-u.ac.jp}
\author{Hajime \textsc{Baba}}
\affil{Center for Planning and Information Systems,
  Institute of Space and Astronautical Science,\\
  Sagamihara, Kanagawa 229-8510}
\email{baba@plain.isas.ac.jp}
\author{Katsura \textsc{Matsumoto}}
\affil{Graduate School of Natural Science and Technology, Okayama University,
       Okayama 700-8530}
\email{katsura@cc.okayama-u.ac.jp}
\author{Taichi \textsc{Kato}}
\affil{Department of Astronomy, Faculty of Science, Kyoto University,
Sakyo-ku, Kyoto 606-8502}


\KeyWords{accretion, accretion disks
          --- stars: novae, cataclysmic variables
          --- stars: dwarf novae
          --- stars: individual (DM Lyr)}

\maketitle

\begin{abstract}

 We carried out time-resolved $V$-band photometry of DM Lyr during long
 outbursts in 1996 July and in 1996 February-March at Ouda Station, Kyoto
 University and at Osaka Kyoiku University.  Since superhumps were
 clearly detected in the light curves, DM Lyr was first identified with
 an SU UMa-type dwarf nova.  The superhump period is 0.0673(2) d, and
 the superhump excess is 2.8(3) \%.  The duration of the superoutburst,
 the outburst amplitude, the decline rate in the plateau phase, and the
 superhump excess were typical values for a usual SU UMa star.
 According to visual and CCD observations reported to VSNET, this star
 has experienced a dramatic change of the outburst pattern from a
 superoutburst phase to a normal outburst phase.  There may exist
 mechanisms to decrease the number of the normal outburst between two
 successive superoutbursts and to elongate the recurrence cycle of the
 superoutburst.

\end{abstract}

\section{Introduction}

Cataclysmic variable stars (CVs) are close binary systems of a white
dwarf (primary star) and a late-type main-sequence star (secondary star)
filling its Roche-lobe (for a review, e.g. \cite{war95book}).  The
surface gas of the secondary is transfered for the primary star, and an
accretion disk is formed around the primary star.  Some CVs
quasi-periodically show sudden brightenings, namely, outbursts.  This
behavior is generally explained by the thermally unstable accretion
disk in the disk instability model (for a review, \cite{osa96review}).
Such CVs are called dwarf novae.

Dwarf novae are classified into three basic subclasses, which are SS Cyg
stars showing (normal) outbursts, Z Cam stars showing normal outbursts
and standstills, and SU UMa stars showing normal outbursts and
superoutbursts (see Ch. 3 in \cite{war95book}).  The basic physics
differentiating these three subclasses is currently understood in the
scheme of the thermal-tidal disk-instability theory mentioned above.
Thorough exploration to explain a variety of behavior of CVs, however,
is being continued.

DM Lyr is a poorly studied dwarf nova.  This star was discovered as a
nova or a U Gem-type variable star by \citet{hof29dmlyr}, who originally
designated it as 250.1929.  He also reported a long outburst which
lasted at least 12 days.  \citet{hof30dmlyr} gave a finding chart.
Based on records of 6 dwarf nova-type outbursts, \citet{pet60dmlyr}
estimated the outburst properties of DM Lyr, which were the recurrence
cycle of outburst of over 100 d, and the magnitude range of 13.7--17.5:.
\citet{BruchCVatlas}, nevertheless, listed DM Lyr as an unidentified
dwarf novae.  \citet{DownesCVatlas1} then correctly re-identified DM
Lyr.  Very recently, \citet{tho03kxaqlftcampucmav660herdmlyr} obtained
spectra having singly-peaked emission lines indicative of a low
inclination, and measured the orbital period (\Porb) to be 0.06546(6) d.

This object is identified with 1RXS J185845.1+301548 \citep{ROSATRXP},
while no counterpart is found in the 2 micron All Sky Survey
\citep{hoa02CV2MASS}.  The X-ray hardness ratios of the object are
compatible with those of dwarf novae.

\begin{table*}
\caption{Log of the observations.}\label{tab:log}
\begin{center}
\begin{tabular}{clccccc}
\hline\hline
\multicolumn{3}{c}{Date (UT)} & N & Exposure & Mean $V$ & Site$^*$ \\
     &       &                &   & time (s) &     Mag. &  \\
\hline
1996 & July     & 15.525 -- 15.752 & 225 & 60  & 14.49 & OS \\
     &          & 16.536 -- 16.752 & 253 & 60  & 14.66 & OS \\
     &          & 17.574 -- 17.703 & 153 & 60  & 14.79 & OS \\
     &          & 17.606 -- 17.769 &  50 & 120 & 14.78 & OKU \\
     &          & 18.584 -- 18.749 &  46 & 150 & 14.97 & OKU \\
     &          & 21.558 -- 21.566 &   6 & 120 & 17.25 & OS \\
     &          & 24.540 -- 24.543 &   4 &  90 & 17.52 & OS \\
     &          & 29.636 -- 29.641 &   5 &  90 & 18.42 & OS \\
1997 & March    &  1.754 --  1.852 & 112 & 60  & 14.28 & OS \\
     &          &  4.790 --  4.863 &  86 & 60  & 14.55 & OS \\
     &          &  5.856 --  5.857 &   3 & 60  & 14.81 & OS \\
     &          &  7.847 --  7.861 &  15 & 60  & 15.21 & OS \\
     &          &  8.848 --  8.852 &   5 & 60  & 16.46 & OS \\
\hline
\multicolumn{7}{l}{$^*$ OS represents Ouda Station, and OKU
 represents Osaka Kyoiku University.}
\end{tabular}
\end{center}
\end{table*}

Based on the long outburst reported by \citet{hof29dmlyr}, we had kept
DM Lyr in mind as a good candidate of an SU UMa-type dwarf nova, and had
watched our chance to reveal its nature.  Under this situation, we
started time-resolved photometry, following the VSNET\footnote{see
htpp://vsnet.kusastro.kyoto-u.ac.jp/vsnet/} report of an outburst caught
at 1996 July 7.040 (UT) by G. Poyner (vsnet-obs 3060) and subsequent
confirmations.  After this outburst, we again carried out photometric
observations of DM Lyr during the outburst in 1997
February-March.  Tentative results were listed in table 1 in
\citet{nog97sxlmi}.  We here report the details of the observations, and
discuss the nature of DM Lyr.

\section{Observation}

\begin{figure}
 \begin{center}
  \FigureFile(88mm,115mm){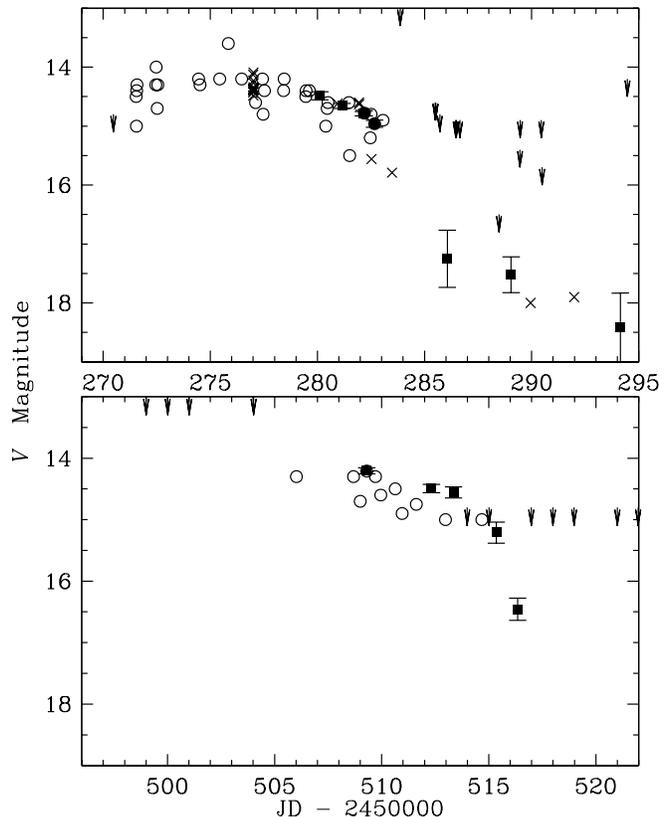}
 \end{center}
 \caption{Outbursts in 1996 July and 1997 February-March.  The open
 circles and crosses are visual observations reported to VSNET and our
 unfiltered CCD observations (25cm Schmidt-Cassegrain telescope + ST-7),
 respectively.  The filled squares represent our observations.  The
 lower arrows indicate the upper limits.
 }
 \label{fig:longlc}
\end{figure}

We performed the observations at the Ouda Station (OS), Kyoto
University, and at Osaka Kyoiku University (OKU).  At OS, a 60-cm
reflector (focal length=4.8 m) and a CCD camera (Thomson TH~7882, 576
$\times$ 384 pixels) attached to the Cassegrain focus were used (for
more information of the instruments, see Ohtani et al. 1992).  At OKU,
we used a 51-cm reflector (focal length=6.0 m) and a CCD camera
(Astromed EEV 88200, 1152 $\times$ 790 pixels).  The on-chip 2 $\times$
2 binning mode was selected to reduce the read-out and saving dead time.
Johnson {\it V}-band interference filters were adopted.  Table
\ref{tab:log} gives the journal of the observation.

After standard de-biasing and flat fielding, the frames obtained at OS on
1996 July 15, 16, and 17, and on 1997 March 1, 4, 5, 7, and 8 were
processed by a microcomputer-based aperture photometry package, and
those obtained on 1996 July 21, 24, 29 were reduced by a PSF
photometry package.  Both packages were developed by one of the authors
(TK).  There is no systematic difference larger than 0.03 mag between
results by aperture photometry and those by PSF photometry.  In the
similar way, the OKU frames were reduced by the aperture photometry,
using the IRAF package\footnote{IRAF is distributed by the National
Optical Astronomy Observatories for Research in Astronomy, Inc. under
cooperative agreement with the National Science Foundation.}.

The magnitude of DM Lyr was measured, relative to the local standard star
GSC 2639.2575 ($V=12.47$ in the VSNET
chart\footnote{ftp://vsnet.kusastro.kyoto-u.ac.jp/pub/vsnet/charts/DM\_Lyr.ps})
at OS and GSC 2639.2348 ($V=12.77$ in the VSNET chart) at OKU.  Then, we
subtracted 0.12 mag from the OKU data to smoothly connect the OKU data
with the OS data.  Using local comparison stars, constancy of the
standard star during our runs was confirmed within 0.06 mag, and the
nominal 1-$\sigma$ error for each point was estimated. Heliocentric
corrections to observation times were applied before the following
analysis.

\section{Result}

\begin{figure}
 \begin{center}
   \FigureFile(88mm,115mm){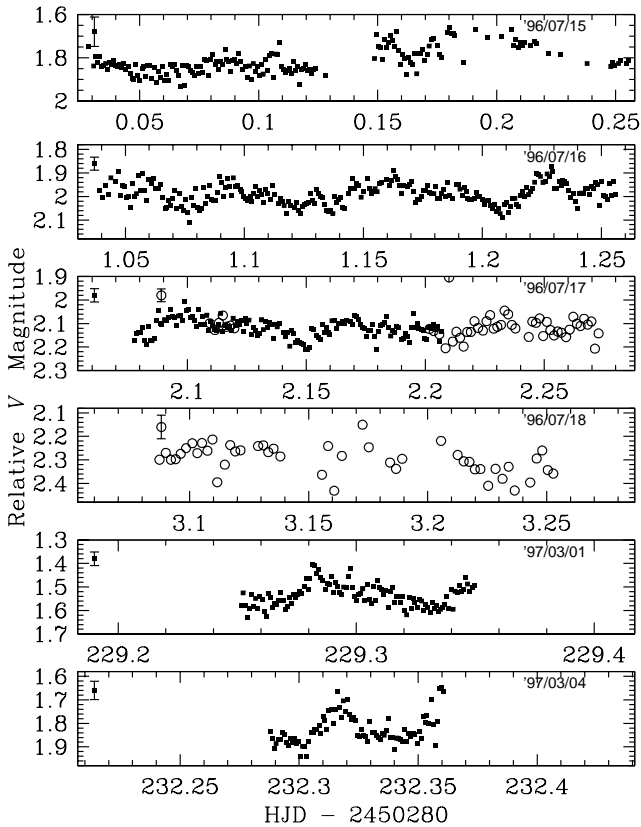}
 \end{center}
 \caption{Short-term light curves with long coverages.  The filled
 squares represent the OS data, and the OKU data are denoted by the open
 circles.  The typical 1-sigma error of each point is indicated at the
 upper-left corner of each panel.  The magnitudes are given relative to
 GSC 2639.2575.  Prominent superhumps are seen, indicative of the SU UMa
 nature of DM Lyr.
 }
 \label{fig:shortlc}
\end{figure}

The over-all light curves of the outbursts in 1996 July and 1997
February-March are shown in figure \ref{fig:longlc}.  The 1996 outburst
was caught on July 7.040 (UT, JD 2450271.540), and lasted at least until
July 19 (JD 2450283), means that the duration was 12 days or some more,
followed by a gradual decline for several days after the rapid decline
from the outburst.  Our observations were performed in the late phase of
this long outburst.  The enlarged light curves of our data are depicted
in figure \ref{fig:shortlc}.  In these light curves, we succeeded in
detecting the clear superhumps with an amplitude of $\sim$0.12 mag on
1996 July 16 and 17, although these oscillations were smeared by the
relatively large errors on July 15 and 18.  This observation revealed
that DM Lyr is really an SU UMa-type dwarf nova and the current outburst
is a superoutburst.

\begin{figure}
 \begin{center}
  \FigureFile(88mm,115mm){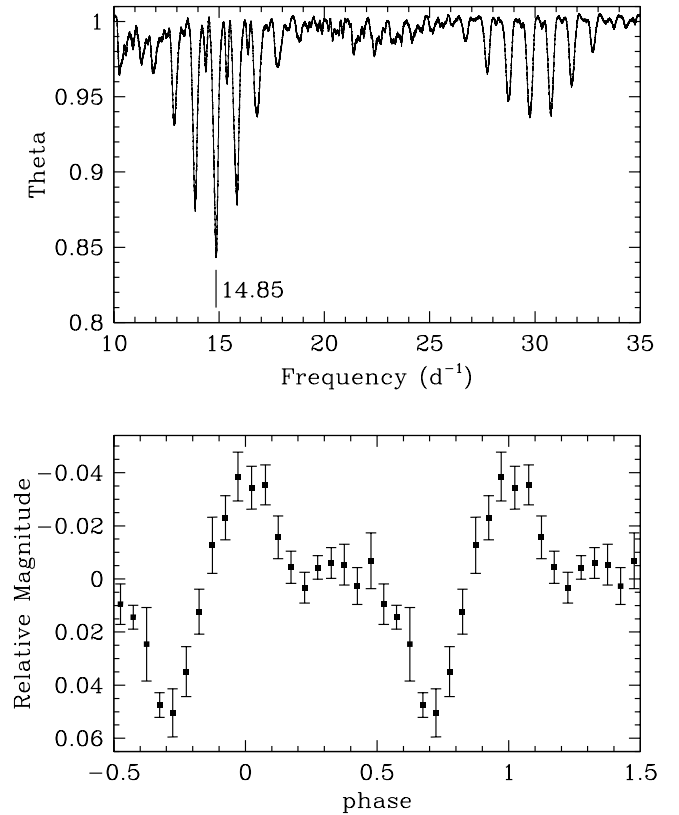}
 \end{center}
 \caption{(upper panel) Theta diagram of the PDM period analysis on the
 data during the 1996 superoutburst.  The superhump period is 0.0673(2)
 d (= 14.85 d$^{-1}$).  (lower panel) The average light curve of the
 superhumps obtained by folding the data by the period of 0.0673 d.
 }
 \label{fig:pdm}
\end{figure}

The linear decline trend between July 15 and 18 was 0.15 mag d$^{-1}$,
which is a typical, or a little bit large value during the plateau phase
in SU UMa stars.  We performed a period analysis of the phase dispersion
minimization (PDM) method \citep{PDM} on the data after subtraction of
the decline trend.  The resultant theta diagram is shown in the upper
panel of figure \ref{fig:pdm}.  The superhump period (\Psh) is securely
determined to be 0.0673(2) d, and the lower panel of figure
\ref{fig:pdm} exhibits the light curve folded by this period.  The
superhumps did not have the textbook shape of a rapid rise and a slow
decline, but had a secondary maximum around the phase of 0.4.  This
secondary maximum may be a hint of late superhumps which grows in the
very late phase of the superoutburst (\cite{hes92lateSH}, and references
therein).

\begin{table}
\caption{The superhump-maximum timings during the 1996 superoutburst.}\label{tab:max}
\begin{center}
\begin{tabular}{r r c r r}
\hline\hline
$E$ & HJD$^*$ & & $E$ & HJD$^*$ \\
\hline
 1 & 0.088 & & 18 & 1.229 \\
 2 & 0.156 & & 31 & 2.099 \\
16 & 1.095 & & 32 & 2.171 \\
17 & 1.160 & & 33 & 2.234 \\
\hline
\multicolumn{5}{l}{$^*$ HJD - 2450280.}\\
\end{tabular}
\end{center}
\end{table}

\begin{figure}
 \begin{center}
  \FigureFile(88mm,115mm){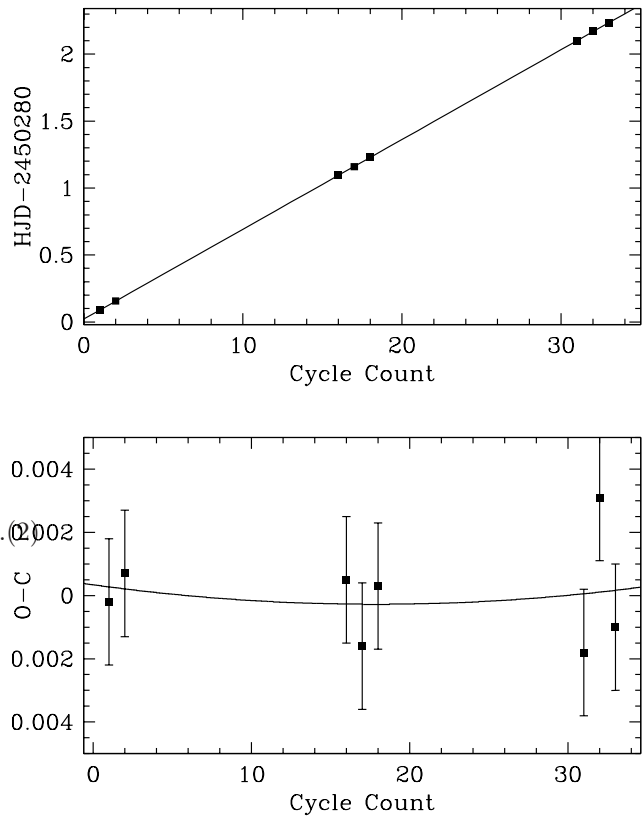}
 \end{center}
 \caption{(upper panel) Superhump maximum timings during the 1996
 superoutburst and its linear regression (equation (1)).  The error bar
 is hidden by the filled marks.  (lower panel) $O-C$ diagram  of the
 superhump maximum timings.  The curve represents the quadratic
 polynomial (equation (2)) obtained by fitting the $O-C$.
 }
 \label{fig:max}
\end{figure}

We measured the maximum timings of these superhumps by eye to check
variation of the superhump period.  Table \ref{tab:max} summarizes the
results.  The typical error of the timings is 0.002 d.  The cycle count
$E$ is adopted to set $E=1$ at the first maximum timing.  The linear
regression gives an equation (figure \ref{fig:max}):
\begin{equation}
HJD = 2450280.0212(11) + 0.06709(5)E.
\end{equation}
The superhump period deduced here is in accordance with the one obtained
by the PDM period analysis within the error.  By fitting the deviations
of the observed timings from the calculation, we obtain a quadratic
polynomial:
\begin{equation}
O-C = 0.0003(15) - 0.7(2.0)\times10^{-4}E + 1.9(5.8)\times10^{-6}E^2.
\end{equation}
The change rate derived from the quadratic term is $\dot{P}_{\rm
sh}/P_{\rm sh} = 5.7(17.2)\times 10^{-5}$, which covers the almost whole
range of distribution of the \Psh\ change rate of SU UMa stars
(see \cite{kat01hvvir}).  Consequently, the present data were not
sufficient to accurately derive the change rate, and even to judge the
sign of the rate.

\begin{figure}
 \begin{center}
  \FigureFile(88mm,115mm){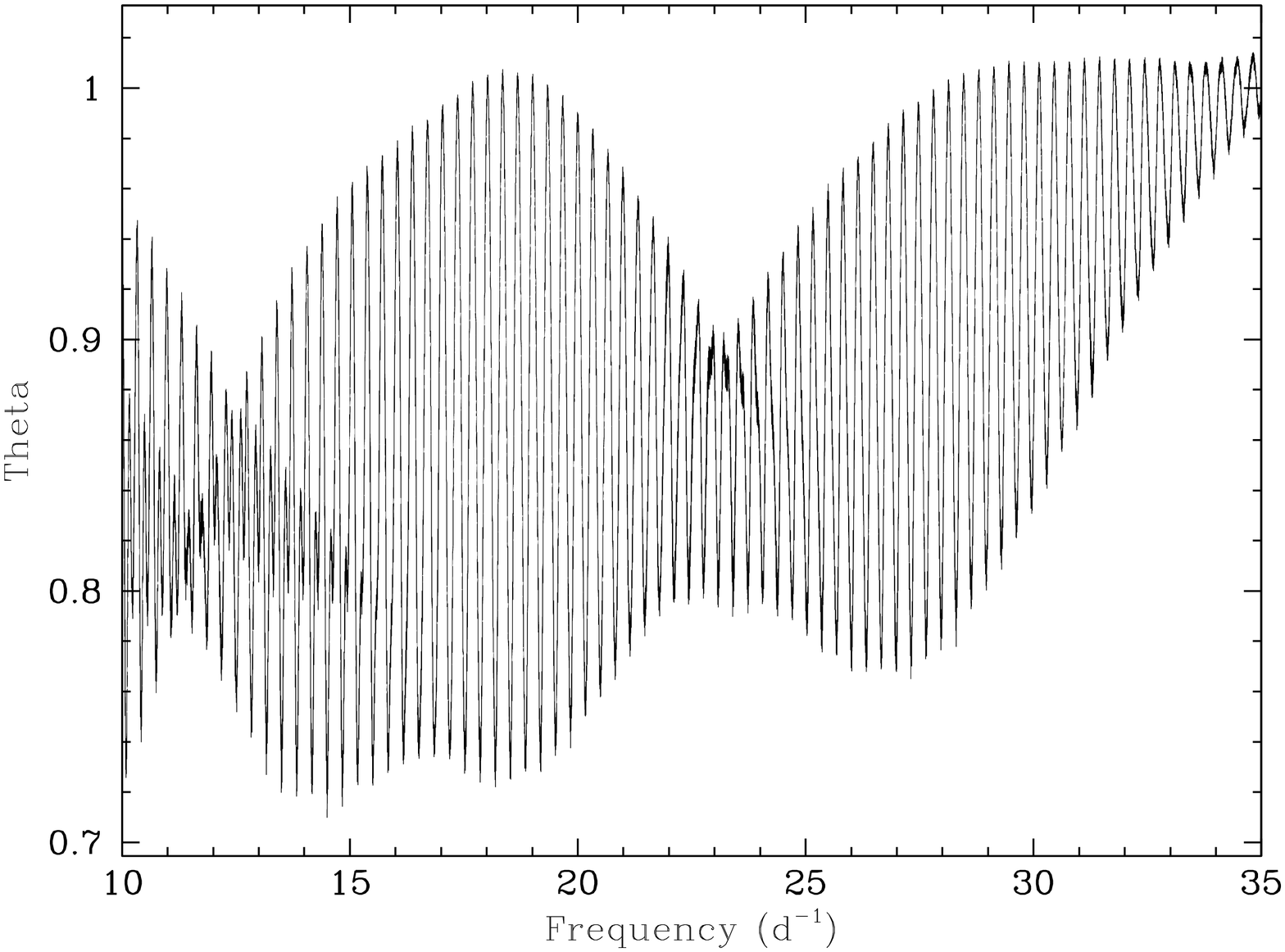}
 \end{center}
 \caption{Theta diagram of the PDM period analysis on the data during
 the 1997 superoutburst.  Due to the short coverages in conjunction with
 the relatively large errors, clear periodicity could not be detected.
 }
 \label{fig:pdm2}
\end{figure}

During the 1997 outburst, we again made time resolved photometry in the
late phase (figure \ref{fig:longlc}).  Superhumps were observed also in
this outburst (figure \ref{fig:shortlc}), which indicates that this is
the second superoutburst observed in DM Lyr, apart from the first
superoutburst by $\sim$230 d.  After subtracting the linear decline trend
of 0.10 mag d$^{-1}$ from the data obtained on 1997 March 1 and 4, we
made a PDM period analysis.  Apparent evidence of periodicity, however,
was not found because of the insufficient coverages, although figure
\ref{fig:pdm2} shows peaks with very small significance around the
superhump frequency.

We took power spectra of each-night data to search for quasi-periodic
oscillations (QPOs), but no signal was detected.

\section{Discussion}

As noted in section 1, \citet{tho03kxaqlftcampucmav660herdmlyr} measured
the orbital period of 0.06546(6) d.  The fractional superhump excess
($\epsilon=$(\Psh$-$\Porb)/\Porb) in DM Lyr is then 2.8(3) \%.
Generally, the superhump period decreases with the superoutburst
evolving in SU UMa stars with a similar orbital period (see
\cite{kat01hvvir}), although the present data were not sufficient to
derive a change of \Psh.  If the superhumps were observed from the
earlier phase in the 1996 superoutburst, the superhump excess was
expected to be a little larger.  The superhump excess of 2.8 \% or a
little larger is, however, just on the known \Porb-$\epsilon$
correlation (e.g. \cite{pat98evolution}).  In this meaning, DM Lyr is a
normal SU UMa-type dwarf nova.

The duration of the 1996 superoutburst was 13$(\pm$1) d, while that of
the 1997 superoutburst was constrained only as being longer than 9 days.
This duration is a typical value for an SU UMa star.

The visual magnitudes reported to VSNET seems to include a rather large
error (see figure \ref{fig:longlc}), probably because of lack of a
reliable sequence of comparison stars' magnitudes and the effect of the
close companion star.  The outburst amplitude is, however, estimated to
be 4.3($\pm$0.5) mag in figure \ref{fig:longlc}, which is supported by
the fact that the General Catalog of Variable Stars \citep{GCVS} lists
13.6-18.0p as the maximum and minimum photographic magnitudes.  This
outburst amplitude is also a typical value for an SU UMa star.

Table \ref{tab:ob} lists all outbursts reported to VSNET so
far\footnote{freely available via the VSNET data browser,
http://vsnet.kusastro.kyoto-u.ac.jp/vsnet/etc/searchobs.html}.  We can
see that the first four superoutbursts almost regularly occurred with a
recurrence times of 229 d, 235 d, and 275 d.  These recurrence intervals
are normal, or a little small values for an SU UMa star with the orbital
period of $\sim$0.067 d (see \cite{nog97sxlmi}).  However, it should be
pointed out that only one normal outburst was caught during this period.
\citet{war95suuma} derived a correlation between the mean recurrence
cycles for normal outbursts ($T_n$) and for superoutbursts (supercycle,
$T_s$).  This correlation expects $T_n$ $\sim$ 46 d for $T_s$ = 250 d.

In contrast, since 1997 November, only one superoutburst in 2000
August was caught for about 5 years.  Instead, many normal outbursts
were repeatedly found with intervals of 14-238 days.  DM Lyr has been
rather closely monitored these years, except in winter when DM Lyr is
not observable.  Therefore, it is unlikely that many superoutbursts have
been missed.  If we assume $T_s$ to be $\sim$950 d, which is the period
between the 1997 November superoutburst and the 2000 July superoutburst,
no normal outburst is expected according to the $T_n$-$T_s$ relation
mentioned above, like WZ Sge-type stars or related objects (see
\cite{kat01hvvir}).  Although normal outbursts may have more easily
escaped from the eye because of the faintness and the short duration (table
\ref{tab:ob}), the frequency of the normal outburst clearly changed between
until 1997 and since 1998.  This change was coincident with the
transition between the {\it high/low states} regarding superoutburst.
In other words, DM Lyr has two states: 1) the superoutburst phase when
the superoutburst happens with a supercycle of $\sim$250 d, but DM
Lyr experience few normal outbursts, and 2) the normal-outburst phase
when the normal outbursts occurs with a recurrence cycle of a few tens-a
few hundreds of days, but the supercycle is $\sim$1000 d.  DM Lyr
sharply showed the transition between these two states.

\begin{table}
\caption{Previous outbursts.}\label{tab:ob}
\begin{center}
\begin{tabular}{llrlrll}
\hline\hline
\multicolumn{3}{c}{Date$^*$} & $V_{max}$ & D$^\dagger$ & Type$^\ddagger$ & Comment \\
\hline
1995 & Nov. & 20 & 13.9  & $>$8  & S \\
1996 & Jul. & 07 & 13.6  & $>$12 & S \\
1996 & Oct. & 15 & 15.2: & 1?    & N? & Single Obs. \\
1997 & Feb. & 26 & 14.2  & $>$10 & S \\
1997 & May  & 25 & 14.9  & 1?    & N \\
1997 & Nov. & 29 & 14.1  & $>$7  & S \\
1998 & Mar. & 10 & 14.8  & $<$3  & N \\
1998 & Jul. & 13 & 14.5  & 3     & N \\
1998 & Aug. & 26 & 15.8  & 1?    & N \\
1998 & Nov. & 09 & 15.2: & 1?    & N? & Single Obs. \\
1998 & Dec. & 08 & 15.0  & 2?    & N \\
1999 & Aug. & 20 & 15.5  & $<$2  & N & Single Obs. \\
1999 & Sep. & 28 & 14.5  & 2     & N \\
2000 & May  & 24 & 15.7: & 1?    & N? & Single Obs. \\
2000 & Jun. & 27 & 15.5: & 1?    & N? \\
2000 & Jul. & 11 & 14.1  & $>$8  & S \\
2000 & Aug. & 25 & 15.5  & 1?    & N \\
2001 & Jul. & 4  & 15.1: & 1?    & N? & Single Obs. \\
2001 & Aug. & 1  & 15.7: & 1?    & N? & Single Obs. \\
2001 & Aug. & 15 & 16.0: & 1?    & N? & Single Obs. \\
2001 & Sep. & 4  & 14.5  & 2?    & N \\
2002 & Feb. & 16 & 16.0: & 1?    & N? & Single obs. \\
2002 & May  & 11 & 15.1  & 3?    & N \\
2002 & Jun. & 3  & 15.7: & 1?    & N? & Single obs. \\
2002 & Jul. & 14 & 15.5  & 2     & N \\
2002 & Aug. & 15 & 14.8  & 3?    & N \\
2002 & Sep. &  8 & 15.5  & 1?    & N  & Single obs. \\
\hline
\multicolumn{7}{l}{$^*$ The discovery date.}\\
\multicolumn{7}{l}{$^\dagger$ Duration of the outburst in a unit of day.}\\
\multicolumn{7}{l}{$^\ddagger$ N: normal outburst, S: superoutburst}
\end{tabular}
\end{center}
\end{table}

The disk instability theory predicts that both of $T_n$ and $T_s$ are
tightly related to the mass transfer rate from the secondary star
($\dot{M}$).  In case of normal SU UMa stars, $T_n$ and $T_s$ is
expected to be roughly proportional to $\dot{M}^{-2}$ and $\dot{M}$,
respectively (see section 5.4 in \cite{osa96review}).  Decrease of the
mass transfer rate is thus predicted to lead to increase of both $T_n$
and $T_s$ and to decrease of the number of the normal outburst during
one supercycle.  This trend well agrees with the empirical $T_n$-$T_s$
relation, although other parameters, such as, the orbital period, masses
of the components, and so on, are neglected in this qualitative
discussion.  The phase transition from the superoutburst phase to the
normal outburst phase in DM Lyr presented here completely opposes this
theoretical and empirical relation in that increase of the supercycle is
accompanied with increase of the number of the normal outburst in one
$T_s$.

Changes of the outburst patterns in SU UMa stars have been reported in
recent years, e.g. in DI UMa \citep{fri99diuma}, SU UMa
(\cite{ros00suuma}; \cite{kat02suuma}), V1113 Cyg \citep{kat01v1113cyg},
and V503 Cyg \citep{kat02v503cyg}.  The behavior of V503 Cyg among these
is closest to the present case. \citet{kat02v503cyg} reported the
dramatic decrease of the number of the normal outburst in one $T_s$, and
proposed existence of a mechanism to decrease or even quench the normal
outburst.  The most important difference between the behavior of V503
Cyg and that of DM Lyr is constancy of $T_s$.  $T_s$ in V503 Cyg has
been almost constant, regardless of decrease of the number of the normal
outburst between two successive superoutbursts.  In DM Lyr, however,
$T_s$ became longer at the same time when normal outbursts came to arise
with shorter $T_n$ of a few tens-a few hundreds of days.

Since the set of $T_s$ in the superoutburst phase and $T_n$ in the
normal outburst phase almost satisfies the empirical $T_n$-$T_s$, the
mass transfer rate may be constant also in DM Lyr, and a mechanism to
decrease the number of the normal outburst had worked during the
superoutburst phase.  Then, a mechanism to elongate $T_s$ might start
working in the normal outburst phase.  To check the stability of the
mass transfer rate, it is important to measure the quiescence magnitude
and the amplitude of the orbital hump during the current normal outburst
phase.  In addition, closer monitoring is needed to confirm the
existence of the future superoutburst phase and to avoid to miss normal
outbursts then.  In SU UMa, \citet{kat02suuma} reported their superhump
detection during a faint outburst (or a minor brightening) which arose
in an anomalously outbursting state, and proposed that an long-lasting,
tidally unstable state following the preceding superoutburst may
suppress normal outbursts.  We should try to make the monitoring going
as deep as possible and to start photometric observations soon on
finding a brightening.  DM Lyr may be a key object to lead to find a new
mechanism to control the state of the accretion disk.

DM Lyr in the superoutburst phase seems a perfect twin of another
peculiar SU UMa-type dwarf nova, V844 Her.  This object show
superoutbursts with $T_s$ of 220-290 d, but no normal outburst has been
caught (see \cite{kat00v844her}, and visual observations available via
the VSNET data browser)\footnote{Very recently, the first-ever normal
outburst was caught in 2002 October 23-25 (see vsnet-campaign-dn 2933)}.
The differences between DM Lyr and V844 Her are that the outburst
amplitude of V844 Her ($\sim$5.7 mag) is larger than that of DM Lyr, and
that the orbital period of V844 Her is near the period minimum (\Porb
= 0.054643(7) d, \cite{tho02gwlibv844herdiuma}).  V844 Her
will possibly change the outburst pattern in future like DM Lyr, and
deserves intensive watch.

\vskip 3mm

We deeply thank amateur observers who have been reporting their valuable
observations to VSNET.  The authors are grateful for the anonymous
referee for the kind comments.

\end{document}